\pdfoutput=1
\documentclass{article}
\usepackage{url,hyperref}
\usepackage{graphicx}
\usepackage{color}

\usepackage[T2A]{fontenc}
\usepackage[utf8]{inputenc}

\newcommand{\normconst}[1]{H(#1)}
\newcommand{\bigo}[1]{O\left(#1\right)}

\title{A Simple Derivation of the Heap's Law from the Generalized Zipf's Law}
\date{November, 2017}
\author{Leonid Boytsov}
\begin{document}
\maketitle
\begin{abstract}
I reproduce a rather simple formal derivation of the Heaps' law
from the generalized Zipf's law,
which I previously published in Russian~\cite{boytsov2003}.
\end{abstract}
\section{Introduction}
There are two well-known regularities in natural language texts,
which are known as Zipf's and Heaps' laws.
According to the original Zipf's law,
a probability of encountering the $i$-th most frequent word $w_i$
is inversely proportional to the word's rank $i$:
$$
p_i = O(1/i).
$$
This law---which was not actually discovered by Zipf \cite{DBLP:conf/conll/Powers98}---is not 
applicable to arbitrarily large texts.
The obvious reason is that the sum of inverse ranks does not converge to a finite number.
A slightly generalized variant, henceforth \textbf{generalized} Zipf’s law,
is likely a more accurate text model:
$$
p_i = O(1/i^{\alpha}),$$
where $\alpha > 1$.

Heaps' law \cite{Heaps1978}---also discovered by Herdan \cite{Egghe:2007:UHL:1231003.1231008}---approximates the number of unique words in the text of length $n$:
$$
|\cup_{i=1}^n \{ w_i \} | = O(n^\beta),$$ where $\beta <1$.
Heaps' law says that the number of unique words grows roughly sub-linearly as a power function of the total number of words (with the exponent strictly smaller than one).

Somewhat surprisingly, Baeza-Yates and G. Navarro \cite{DBLP:journals/jasis/Baeza-YatesN00}
argued (although a bit informally) that constants in Heaps' and Zipf's laws are reciprocal numbers:
$$
\alpha \approx 1/\beta.
$$
They also verified this empirically.

This work inspired several people (including yours truly) to formally derive Heaps' law from the generalized Zipf's law (all derivations seem to have relied on a text generation process where words are sampled independently).
It is hard to tell who did it earlier. 
In particular, I have a recollection that Amir Dembo (from Stanford) produced an analgous derivation (not relying on the property of the Gamma function), but, apparently, he did not publish his result.
Leijenhorst and Weide published a more general result 
(their derivation starts from Zipf-Mandelbrot distribution rather than from generalized Zipf)
in 2005 \cite{VANLEIJENHORST2005263}.
My own proof was published in Russian in 2003 \cite{boytsov2003}.
Here, I reproduce it for completeness. I tried to keep it as simple as possible:
some of the more formal argument is given in footnotes.

\section{Formal Derivation}
As a reminder we assume that the text is created by a random process where words are sampled independently
from an infinite vocabulary. 
This is not the most realistic assumption, however, it is not clear how one can incorporate
word dependencies into the proof.
The probability of sampling the $i$-th most frequent word
is defined by the generalized Zipf's law, i.e.,

$$
 p_i = \frac{1}{\normconst{\alpha} \cdot i^{\alpha}}, 
\eqno(*)
$$
where $\alpha > 1$ and 
$\normconst{\alpha} = \sum_{i=1}^{\infty} {1 / i^{\alpha}}$
is a normalizing constant.

The number of unique words in the text is also a random variable $X$,
which can be represented as an infinite sum of random variables $X_i$.
Note that $X_i$ is equal to one if the text contains at least one word $w_i$ and is zero otherwise.
The objective of this proof is to estimate the expected number of unique words $E X$:
$$
E X =  E \left(\sum_{i=1}^{\infty} X_i\right)
$$
The proof only cares about an asymptotic behavior of $E X$ with respect to the total
number of text words $n$, i.e., all the derivations are big-O estimates.

Because words are sampled randomly and independently,
a probability of \emph{not} selecting word $w_i$ after $n$ trials is equal to $(1-p_i)^n$.
Hence, $X_i$ has the 
Bernoulli distribution with the success probability
$$
 p(X_i = 1) = 1 - (1-p_i)^n,
$$
where $p_i$ is a probability of word occurrence according to the generalized Zipf's law
given by Eq.~(*).
Therefore, we can rewrite the expected number of unique words as follows:
$$
 E X = \sum_{i=1}^{\infty} 1 - (1-p_i)^n
=
\sum_{i=1}^{\infty} 1 - \left( 1 - {1 \over {\normconst{\alpha} i^\alpha}}\right)^n
$$

What can we say about this series in general and about the summation term $1 - (1-p_i)^n$ in particular?
\begin{itemize}
\item Because $0 < 1 - (1-p_i)^n < n \cdot p_i$ and $\{p_i\}$ is a convergent series,
our series converges.\footnote{The upper
bound for the series term follows
from $1-q^n = (1-q)(q^{n-1} + q^{n-2} + \ldots +1)$, which is upper bounded
by $(1-q) \cdot n$ for $0 < q < 1$ and positive $n$. }
\item The summation term can be interpreted as a real valued function of the variable $i$.
The value of this function decreases monotonically with $i$. 
The function is positive for $i \ge 0$ 
and is upper bounded by one.\footnote{$p_i$ decreases with $i$; $1-p_i$ increases with $i$;
$(1-p_i)^n$ increases with $i$; $1 - (1-p_i)^n$ decreases with $i$.}
\end{itemize}
Thanks to these properties, we can replace 
the sum of the series with the following big-O equivalent integral from $1$ to $\infty$:\footnote{
Using monotonicity it is easy to show that the integral from 1 to $\infty$ is smaller than 
the sum of the series, but the integral from 0 to $\infty$ is larger than
the sum of the series. The difference between two integral values is less than one.}
$$
 \int_{1}^{\infty} 1 - \left( 1 - {1 \over {\normconst{\alpha} x^\alpha}}\right)^n dx
\eqno(**)
$$
Using the variable substitution
 $y = x \normconst{\alpha} ^{1\over \alpha}$,
we rewrite (**) as follows:

$$
 {1 \over { \normconst{\alpha} } } \int\limits_{ \normconst{\alpha} ^{1\over \alpha}}^{\infty} 1 - \left( 1 - {1 \over y^\alpha}\right)^n dy 
$$

Because the integrand is positive and upper bounded by one,
the value of the integral for the segment
$[0, \normconst{\alpha}  ^{1\over \alpha}] $ 
is a constant with respect to $n$.
$\normconst{\alpha}$ is a constant as well.
Therefore, the value of the integral is big-O equivalent to the value of the following integral
which goes from one to infinity:
$$
 \int\limits_{1}^{\infty} 1 - \left( 1 - {1 \over y^\alpha}\right)^n dy 
$$
We further rewrite this by applying the Binomial theorem to the integrand: 
$$
 \int\limits_{1}^{\infty} 1 - \left( 1 - {1 \over y^\alpha}\right)^n dy  =
 \int\limits_{1}^{\infty} \left(1 - \sum_{i=0}^n (-1)^i C^{i}_n {1 \over y^{\alpha i}}\right) dy =
 \int\limits_{1}^{\infty} \left(\sum_{i=1}^n (-1)^i C^{i}_n {1 \over y^{\alpha i}}\right) dy 
$$
Because $\alpha > 1$, every summand in the integrand has absolute convergence.\footnote{This is 
concerned with the convergence of the integral with respect to its infinite upper bound.}
Hence, the integral of the finite sum is equal to the following sum of integrals:
$$
 \sum_{i=1}^n C^{i}_n (-1)^i \int\limits_{1}^{\infty}  {1 \over y^{\alpha i}} dy 
= \sum_{i=1}^n C^{i}_n (-1)^i \left({1\over {i\alpha - 1}} \right)
=
$$
$$
= {1\over \alpha} \sum_{i=1}^n C^{i}_n (-1)^i \left({1\over {i - (1/\alpha) }} \right) =
$$
(because the term for $i=0$ is equal to minus one)
$$
=
 1 +{1\over \alpha} \sum_{i=0}^n C^{i}_n (-1)^i \left({1\over {i - (1/\alpha) }} \right)
\eqno(***)
$$
Using induction one can demonstrate that (also see
\cite[\S 1.2.6, Exercise 48]{DBLP:books/lib/Knuth98}):
$$
 \sum_{i \ge 0 } C^i_n { (-1)^i \over i + x } = { n! \over x (x+1) \ldots (x+n)}
$$
This allows to rewrite Eq.~(***) as follows:
$$
 1 + {{ 1 / \alpha } \cdot \;n! \over { (-1/\alpha) (1-1/\alpha) \ldots ( n - 1/\alpha)  }}
$$
Now, using the formula 
$$
 \Gamma (x) = \lim_{n \rightarrow \infty} 
{n^x n! \over x (x+1)  \ldots (x+n) }
$$
and its corrollary
$$
 { n! \over x (x+1)  \ldots (x+n) } = \bigo{\Gamma(x) \cdot n^{-x}}  
$$
with $x=-1/\alpha$ we obtain that (***) is big-O equivalent
to 
$$\Gamma(-1/\alpha) \cdot n^{1 / \alpha} = \bigo{n^{1/\alpha}}.$$
In other words, the constant $\beta$ is in the Heaps' law is inversely
proportional to the constant $\alpha$ in the generalized Zipf's law.


\begin{thebibliography}{1}

\bibitem{DBLP:journals/jasis/Baeza-YatesN00}
Ricardo~A. Baeza{-}Yates and Gonzalo Navarro.
\newblock Block addressing indices for approximate text retrieval.
\newblock {\em {JASIS}}, 51(1):69--82, 2000.

\bibitem{Egghe:2007:UHL:1231003.1231008}
Leo Egghe.
\newblock Untangling {Herdan}'s {Law} and {Heaps}' {Law}: Mathematical and
  informetric arguments.
\newblock {\em J. Am. Soc. Inf. Sci. Technol.}, 58(5):702--709, March 2007.

\bibitem{Heaps1978}
H.~S. Heaps.
\newblock {\em Information Retrieval: Computational and Theoretical Aspects}.
\newblock Academic Press, Inc., Orlando, FL, USA, 1978.

\bibitem{DBLP:books/lib/Knuth98}
Donald~Ervin Knuth.
\newblock {\em The art of computer programming, Volume {II:} Seminumerical
  Algorithms, 3rd Edition}.
\newblock Addison-Wesley, 1998.

\bibitem{DBLP:conf/conll/Powers98}
David M.~W. Powers.
\newblock Applications and explanations of {Zipf}'s {Law}.
\newblock In {\em Proceedings of the Joint Conference on New Methods in
  Language Processing and Computational Natural Language Learning, NeMLaP/CoNLL
  1998, Macquarie University, Sydney, NSW, Australia, January 11-17, 1998},
  pages 151--160, 1998.

\bibitem{VANLEIJENHORST2005263}
D.C. van Leijenhorst and Th.P. van~der Weide.
\newblock A formal derivation of {Heaps}' {Law}.
\newblock {\em Information Sciences}, 170(2):263 -- 272, 2005.

\bibitem{boytsov2003}
ЛМ Бойцов.
\newblock Синтез системы автоматической
  коррекции, индексации и поиска текстовой
  информации, 2003.

\end{thebibliography}

\end{document}